\documentclass[aps,twocolumn,showpacs,preprintnumbers,pra,amsmath,a4paper,superscriptaddress]{revtex4}
\usepackage{graphicx}
\usepackage[T1]{fontenc}
\usepackage{bm}
\usepackage[colorlinks=false,bookmarks=true,bookmarksnumbered=true]{hyperref}

\newcommand{\ket}[1]{\left| #1\right\rangle}

\newcommand{\avg}[1]{\langle #1\rangle}
\renewcommand{\o}{\omega}
\newcommand{\LNL}{L_\mathrm{NL}}
\newcommand{\eeqref}[1]{Eq.~(\ref{#1})}

\begin{document}                
\author{Wojciech Wasilewski}
\affiliation{Institute of Physics, Nicolaus Copernicus University, Grudzi{\k a}dzka 5, 87-100 Toru{\'n}, Poland}
\affiliation{Institute of Experimental Physics, Warsaw University, Ho{\.z}a 69, 00-681 Warsaw, Poland}
\author{A. I.~Lvovsky}
\affiliation{Department of Physics and Astronomy, University of Calgary, Calgary, Alberta T2N 1N4, Canada}
\author{Konrad Banaszek}
\affiliation{Institute of Physics, Nicolaus Copernicus University, Grudzi{\k a}dzka 5, 87-100 Toru{\'n}, Poland}
\author{Czes{\l}aw Radzewicz}
\affiliation{Institute of Experimental Physics, Warsaw University, Ho{\.z}a 69, 00-681 Warsaw, Poland}
\date{\today}
\title{Pulsed squeezed light: simultaneous squeezing of multiple modes}%


\hyphenation{cryp-to-gra-phy}
\newcommand{\Name}[1]{#1,}
\newcommand{\Vol}[1]{{\bf #1},}
\newcommand{\Year}[1]{(#1)}
\newcommand{\Page}[1]{#1}
\newcommand{\Book}[1]{{\it #1}}
\newcommand{\Publ}[1]{(#1)}
\newcommand{\Review}[1]{{#1}}
\renewcommand{\And}[1]{and #1}
\begin{abstract}
We analyze the spectral properties of squeezed light produced by means of pulsed, single-pass degenerate parametric
down-conversion. The multimode output of this process can be decomposed into characteristic modes undergoing
independent squeezing evolution akin to the Schmidt decomposition of the biphoton spectrum. The main features of this
decomposition can be understood using a simple analytical model developed in the perturbative regime. In the strong
pumping regime, for which the perturbative approach is not valid, we present a numerical analysis, specializing to the
case of one-dimensional propagation in a beta-barium borate waveguide. Characterization of the squeezing modes
provides us with an insight necessary for optimizing homodyne detection of squeezing. For a weak parametric process,
efficient squeezing is found in a broad range of local oscillator modes, whereas the intense generation regime places
much more stringent conditions on the local oscillator. We point out that without meeting these conditions, the
detected squeezing can actually diminish with the increasing pumping strength, and we expose physical reasons behind
this inefficiency.
\end{abstract}

\pacs{42.50.Dv, 03.67.-a}
\maketitle

\section{Introduction}
Squeezed states of the radiation field are those in which the field fluctuations in one of the quadratures are reduced
below the vacuum noise level. They were among the first nonclassical states of light generated experimentally
\cite{Slusher85,Wu86} in the 1980s, and were considered promising as a way of overcoming the shot-noise precision
restrictions in optical measurements \cite{Grangier87} and enhancing the capacity of communication channels
\cite{YuenShapiro80}. Nowadays, the interest to generate squeezed light is mainly due to its applicability in quantum
information and communication. In this context it is viewed primarily as a means of generating quadrature entanglement
\cite{Ou92}, a valuable resource that can be applied, for example, for quantum teleportation \cite{Furusawa98},
computation \cite{LLoyd99}, cryptography \cite{Ralph00} and manipulation of atomic quantum states \cite{Kuzmich00}.
There has been also a theoretical effort towards developing protocols for the continuous-variable error correction \cite{Braunstein982,Lloyd98}
and entanglement purification \cite{Duan00}. Furthermore, operations that enable entanglement distillation in the
continuous-variable domain have been described \cite{BrowEisePRA03}.


The present paper concentrates on one experimental method for producing optical squeezing, namely an optical parametric
amplifier (OPA), implementing degenerate parametric
down-conversion in a single-pass, pulsed configuration. A strong pump pulse propagating through a
nonlinear crystal generates pairs of down-converted photons. The idealized experimental arrangement would be such
that the photons in each generated pair are completely indistinguishable and contribute to the same spatio-temporal mode of the generated field.
A combination of a
high pump pulse energy and high optical nonlinearity leads to multiple pair creation
events, which can be described in the Heisenberg picture as a linear transformation between the field operators at the input $\hat a_{in}$ and the output $\hat{a}_{out}$ of the amplifier:
\begin{equation}
\hat a_{out} = \hat a_{in}\,\cosh\zeta + \hat a^{\dagger}_{in}\,\sinh\zeta. \label{SqOp}\end{equation} When the
initial state of the field is vacuum, the above equation defines a squeezed state with the degree of squeezing
$\zeta$, corresponding to the mean square quadrature noise scaling factor equal to $e^{-2\zeta}$. A crucial assumption
underlying \eeqref{SqOp}, frequently used in theoretical works, is that all the down-converted photons are emitted to
a single mode described by the annihilation operator $\hat{a}_{out}$. However, this does not correspond to a realistic
experimental situation.

In an experiment, pulsed squeezed light was realized for the first time by Slusher et al. \cite{laporta} and has since
been reproduced and improved by a number of researchers \cite{Kumar, Kim, Anderson, Daly, Smithey}. In all these
experiments, a pulsed laser is first frequency doubled and then undergoes parametric down-conversion in a spatially-
and spectrally-degenerate type I configuration. The generated optical state is then characterized by means of homodyne
detection. In contrast to the continuous-wave regime, the pulsed squeezed mode does not associate with a sideband of
the carrier wave, but is localized within the temporal frame of the pump laser pulse \cite{Sasaki05}. This appealing feature is however
compromised by a complex spectral structure of the generated photon pairs. In single-pass down-conversion, biphotons
are emitted not necessarily into a particular mode, but into a wide range of modes restricted only by the energy-conservation and
phase-matching conditions. The usual absence of a dominant spatio-temporal mode can lead to difficulties when we
attempt to observe quantum interference of the generated squeezed vacuum with other classical or nonclassical optical
states. The negative effects range from efficiency reduction in homodyne detection to extra dark counts in single
photon measurements.

The present paper is dedicated to studying the spectral mode of the pulsed squeezed state. A particular question we aim
to answer is the following: can we define a particular mode in which high-efficiency squeezing is achieved? Are there
methods to engineer this mode and what are the most efficient ways to access it in homodyne measurements? The arguments
given in the previous paragraph may lead one to believe that there is no mode in which high-efficiency squeezing is
present. Surprisingly, we demonstrate just the opposite: we find modes in which pure squeezed states are generated, and
also a whole continuum of their phase-locked superpositions turns out to become efficiently squeezed. In fact, good
squeezing is present in a broad, continuous range of pulsed modes that are temporally synchronized with the pump.

The basic theoretical tool in our analysis will be the Bloch-Messiah reduction of an arbitrary Bogoliubov
transformation of bosonic field operators \cite{BraunsteinSq}. This procedure will allow us to identify a set of
independent squeezing modes whose evolution takes a simple form analogous to \eeqref{SqOp}. Such a decomposition
results directly from the preservation of the commutation relations between the field operators in course of the
parametric evolution. This fact allows us to apply the numerical technique of singular-value decomposition to integral
kernels describing the transformation of the field operators, which provides effective means to analyze a realistic
parametric amplifier. The decomposition into independent squeezers has been given previously by Bennik and Boyd
\cite{Bennink}, who deduced it using intricate operator algebraic methods \cite{Ma1990}.

As a physical example, we analyse one-dimensional propagation in a beta-barium borate (BBO) crystal serving as a model
of a waveguide amplifier. The nonlinearity of a BBO crystal has an instantaneous character \cite{BoydNLO}, as this
medium is transparent from infrared to ultraviolet and we find that the downconversion spectrum is limited by the
dispersion across the signal frequencies. This differs from the example studied in Ref.~\cite{Bennink} which assumed
noninstantaneous second-order nonlinearity leading to limited downconversion spectrum, while the dispersion within the
bandwidth of down-converted frequencies was neglected. Also, we discuss in detail the perturbative regime of weak
conversion efficiency which provides intuitive insights into the structure of squeezing building on previous results
obtained for spectral two-photon wave functions \cite{Keller,GricePRA97,Law00}. We also show that the characteristic
squeezing modes of a parametric amplifier provide a general answer to the question of optimizing the local oscillator
shape in homodyne measurements \cite{MJWernerPRA95,ShapiroJOSAB97}.

This paper is organized as follows. In Sec.~\ref{sec:theory} we give theoretical foundations of the decomposition used
to analyze the structure of squeezing in the parametric process. Then in Sec.~\ref{sec:1pair} we discuss the weak
pumping regime with a low pair creation probability per pulse and show that the squeezing modes are defined primarily
by the material dispersion coefficients. The general case of strong pump fields which needs to be solved numerically
is addressed in Sec.~\ref{sec:intense}. The consequences of the multimode character of the output field for homodyne
detection are studied in detail in Sec.~\ref{sec:homodyne}. Finally, Sec.~\ref{sec:conclusions} concludes the paper.

\section{General model of an OPA}\label{sec:theory}
The subject of our analysis is one-dimensional propagation of an optical field through a nonlinear medium in which
three-wave mixing takes place. For simplicity and specificity, we will focus here on the one-dimensional case that can
be realized in non-linear waveguides \cite{NonlinearWaveguides}, which recently have become available commercially.
The one-dimensional case nevertheless includes all the complexities of multimode propagation in the spectral degree of
freedom, and the generalization of the formalism introduced here to the full three-dimensional case is
straightforward.

If the pump field is strong enough to neglect its depletion as well as fluctuations, the propagation
equations in the Heisenberg picture become linear in the signal field. Following the standard nonlinear optics description of
the evolution of the fields with respect to the distance covered rather than time, we shall adopt the modal
decomposition of the electric field operator in the form 
\begin{equation}
\hat E(z,t)=i\int \frac{d\omega}{2\pi}\sqrt{\frac{\hbar\omega}{2\epsilon_0\sqrt{n(\omega)}}} \,e^{-i\omega
t}\hat{a}(\omega;z)+ \text{H.c.} \label{Eq:EztDecomposition}
\end{equation}
where $\epsilon_0$ is the permittivity of the vacuum, $n(\o)$ is the refractive index of the medium at a given frequency
and polarization direction, $\hat a(\omega;z)$ is the annihilation operator of a monochromatic mode at a given propagation stage
indicated by $z$. We have neglected the dependence of the field on transverse coordinates, as we concentrate here on the one-dimensional
case of a waveguide. The quantization of the classical propagation equations of the
three-wave mixing problem \cite{KlasycznaReferencja}
is done by a formal substitution of classical amplitudes by field operators.
This yields a first-order differential equation for annihilation operators $\hat{a}(\omega;z)$ of the form:
\begin{eqnarray}
\lefteqn{\frac{\partial \hat a(\omega;z)}{\partial z} = i k(\omega) \hat a(\omega;z)}
& & \label{aprop} \\
& &
+ \frac{1}{ \LNL E_0} \int d\omega'\,e^{i k_p(\omega'+\omega)z}E_p(\omega'+\omega)\hat a^\dagger(\omega';z) \nonumber
\end{eqnarray}
where the first term on the right-hand side represents the free propagation of the field in the medium, while the
second one represents nonlinear interaction. In the above expression, $E_p(\omega)$ is the spectral amplitude of the
pump field at the point $z=0$, which is chosen half-way through the waveguide of a total length $L$, $\omega_p$ is the
central frequency of the pump spectrum, $k(\omega)$ and $k_p(\omega)$ are the signal and the pump field wavevectors,
respectively, and we rescaled the spectral amplitude of the pump field by introducing $E_0=\int d\omega E_p(\omega)$.
For all calculations in this paper, $E_p(\omega)$ is assumed real and positive, i.e. the pump pulse is prepared so that in the
middle of the waveguide it is not chirped.

With this choice, the multiplicative constant $\LNL$ is the characteristic length of the nonlinear interaction defined
as
\begin{equation}\label{defLNL}
\frac{1}{\LNL}=\frac{\o_p^2\,d_\mathrm{eff} E_0}{8c^2k(\o_p/2)},
\end{equation}
where $d_\mathrm{eff}$ is the effective nonlinearity coefficient for the given polarization directions in the medium
and $\o_p/2$ is assumed to be the central frequency of the signal field. When introducing $\LNL$, we assumed that the
frequency dependence of $d_\mathrm{eff}$ combined with the normalization factors in the modal decomposition given in
Eq.~(\ref{Eq:EztDecomposition}) yields a constant whose changes over the bandwidth of the signal field can be
neglected. The physical interpretation of $\LNL$ is that when the signal field is restricted to a single frequency mode
and no phase mismatch with the pump field is present, $\LNL$ is the length over which the minimum and the maximum RMS
quadrature noise scales by $e$. This corresponds to the single-mode degree of squeezing introduced in Eq.~(\ref{SqOp})
equal to $\zeta=L/\LNL$. Let us note that $1/\LNL$ is proportional to the pump pulse amplitude, which we will further
refer to as the pumping strength.

It is instructive to recast Eq.~(\ref{aprop}) into a standard form which allows us to compare
different quantum mechanical pictures of the evolution of a system. Explicitly, the development
of the field operators described by Eq.~(\ref{aprop}) can be written as:
\begin{equation}\label{Heisenberg}
\frac{\partial \hat a(\omega;z)}{\partial z}=i[\hat H,\hat a(\omega;z)]
\end{equation}
where the operator $\hat{H}$, formally equivalent to the Hamiltonian governing the evolution of the system, is
given by a sum of two terms
$\hat{H}=\hat{H}_0+\hat{V}(z)$. The first term generates linear propagation:
\begin{equation}
\hat H_0 = -\int d\omega\, k(\omega) \hat a^\dag(\omega) \hat a(\omega) \label{H0}
\end{equation}
while the second term is responsible for the three-wave mixing process:
\begin{eqnarray}
\lefteqn{\hat V(z) = \frac{1}{2\LNL E_0} \int d\omega d\omega' }\label{Ham}
&&\\
&&\times
[i e^{i k_p(\omega'+\omega)z}E_p(\omega'+\omega)\hat a^\dagger(\omega)\hat a^\dagger(\omega')+{\rm H.c.}].  \nonumber
\end{eqnarray}
The evolution of the quantum state of light while propagating through the crystal can in principle be evaluated in the
interaction picture by integrating over $z$ the term $\hat{V}(z)$. This is however
highly complicated because the Hamiltonian (\ref{Ham}) is position-dependent and generally does not commute with itself
at different values of $z$. Consequently, the evolution operator needs to be appropriately ordered in its expansion in
$\hat{V}(z)$. As we discuss in the next section and in the Appendix, this ordering can sometimes be neglected and
decomposition into squeezing eigenmodes can be performed efficiently using the interaction picture. The general case is
however best described using the Heisenberg picture approach.

As the propagation equation given in \eeqref{aprop} is linear in the
field operators, for any given waveguide length $L$ we can find
Green functions $C(\omega,\omega')$ and $S(\omega,\omega')$ that
transform the input field operators $a^{in}(\omega)\equiv
a(\omega;-L/2)$ into output field operators $a^{out}(\omega)\equiv
a(\omega;L/2)$ according to:
\begin{equation}
\hat a^{out}(\omega) = \int d\omega' [C(\omega,\omega')\,\hat a^{in}(\omega') + S(\omega,\omega')\,\hat a^{in\,\dagger}(\omega')].
\label{Bvtransform}
\end{equation}
The above Bogoliubov transformation can be brought into a canonical form by applying the Bloch-Messiah
reduction \cite{BraunsteinSq} which we will now briefly review. The relevant mathematical tool is the singular value
decomposition applied to the Green functions $C(\omega,\omega')$ and $S(\omega,\omega')$. Because the output
annihilation operators must satisfy canonical commutation relations, the functions $C(\omega,\omega')$ and
$S(\omega,\omega')$ are connected through a number of relations. In particular, the singular value
decompositions of both $C(\omega,\omega')$ and $S(\omega,\omega')$ can be represented using a shared set of parameters
and functions. The explicit formulas have the following form:
\begin{eqnarray}
C(\omega,\omega') &=& \sum_{n=0}^\infty \psi^*_n(\omega) \,\cosh\zeta_n \,\phi_n(\omega') \nonumber \\
S(\omega,\omega') &=& \sum_{n=0}^\infty \psi^*_n(\omega) \,\sinh\zeta_n \,\phi^*_n(\omega')
\label{Sdecomp}
\end{eqnarray}
where $\phi_n(\omega)$ and $\psi_n(\omega)$ are two orthonormal bases and $\zeta_n$ are real nonnegative parameters.
The existence of such a joint decomposition allows us to introduce two sets of operators for input and output modes, defined according to
\begin{eqnarray}
\hat b^{in}_n&=&\int d\omega \,\phi_n(\omega) \hat a^{in}(\omega) \nonumber \\
\hat b^{out}_n&=&\int d\omega \,\psi_n(\omega) \hat a^{out}(\omega)\,,\label{bns}
\end{eqnarray}
 where $\hat b^{in}_n$ and $\hat b^{out}_n$ are the corresponding annihilation
operators which satisfy the standard commutation relations. Consequently, we
can recast the general transformation given in \eeqref{Bvtransform} into a simple form of single-mode squeezing
transformations acting in parallel on orthogonal modes:
\begin{equation}
\hat b^{out}_n = \cosh\zeta_n\,\hat b^{in}_n + \sinh\zeta_n\,\hat
b^{in\,\dagger}_n, \label{SqOpm}\end{equation} bringing us back to
the elementary expression given in \eeqref{SqOp}.

\begin{figure}
  \center\includegraphics[scale=0.4]{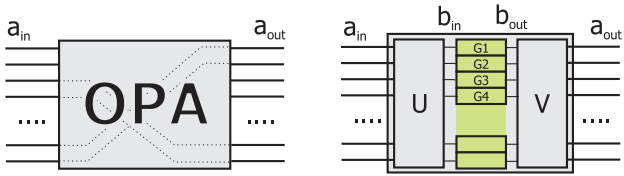}\\
  \caption{Pictorial representation of the Bloch-Messiah theorem.
  Any Bogoliubov transformation of input modes
  $\hat a^{in}_n$ into output modes $\hat a^{out}_n$ can be decomposed into a passive transformation
  $U={\phi_n}$  from modes $\hat a^{in}_n$ into $\hat b^{in}_n$ followed by {\it independent} squeezers
  with gains $G_n$ and another passive transformation $V={\psi^*_n}$ from $\hat b^{out
  }_n$ into $\hat a^{out}_n$.}
  \label{opa:fdecompose}
\end{figure}

Since any linear transformation of the form given in \eeqref{Bvtransform} can be decomposed into orthogonal modes
which undergo independent evolution, the above procedure, summarized graphically in Fig.~\ref{opa:fdecompose}, can be
applied to any combination of processes such as linear optics, optical amplification, and Kerr squeezing. Were we able
to separate modes $\psi_n(\omega)$ spatially, we could turn a realistic parametric amplifier into a generator of
multiple pure squeezed states~\cite{Opatrny}.

Let us note that the procedure of singular value decomposition applied above is formally equivalent to Schmidt
decomposition of a quantum state of a bipartite system. In the next section and in the Appendix, this equivalence will
allow us to connect the general decomposition to previous treatments of parametric down-conversion in the perturbative
regime.

One interesting feature of the squeezing eigenmodes is that for real $E_p$ (i.e. the pump pulse in the middle of the
waveguide is not chirped or its chirp is insignificant),
\begin{equation}
\psi_n(\omega)=\phi^*_n(\omega). \label{conjmodes}
\end{equation}
This is proven by noticing the symmetry of \eeqref{aprop} with respect to the point $z=0$. Indeed, it can be rewritten
as follows:
\begin{eqnarray}
\lefteqn{\frac{\partial \hat a(\omega;z)}{\partial (-z)} = -i k(\omega) \hat a(\omega;z)}
& & \label{apropx} \\
& & - \frac{1}{ \LNL E_0} \int d\omega'\,e^{-i k_p(\omega'+\omega)(-z)}E_p(\omega'+\omega)\hat a^\dagger(\omega';z),
\nonumber
\end{eqnarray}
which can be formally interpreted as reverse propagation of the quantum field from $-z=-L/2$ to $-z=L/2$. Integrating
\eeqref{apropx} between these two points we write, by analogy with \eeqref{Bvtransform}:
\begin{equation}
\hat a^{in}(\omega) = \int d\omega' [\tilde{C}(\omega,\omega')\,\hat a^{out}(\omega) + \tilde{S}(\omega,\omega')\,\hat
a^{\dagger\,out}(\omega)], \label{Bvtransformx}
\end{equation}
where the superscripts $in$ and $out$ refer, as previously defined, to the points $z=-L/2$ and $z=L/2$, respectively.
Comparing Eqs.~(\ref{aprop}) and (\ref{apropx}), we establish a simple relation between the Green functions of direct
and reverse propagation: $\tilde{C}(\omega,\omega')=C^*(\omega,\omega')$ and
$\tilde{S}(\omega,\omega')=-S^*(\omega,\omega')$. Decomposing the latter according to Eq.~(\ref{Sdecomp})
\begin{eqnarray}
\tilde{C}(\omega,\omega') &=& \sum_{n=0}^\infty \psi_n(\omega) \,\cosh\zeta_n \,\phi^*_n(\omega') \nonumber \\
\tilde{S}(\omega,\omega') &=& \sum_{n=0}^\infty \psi_n(\omega) \,\sinh(-\zeta_n) \,\phi_n(\omega'), \label{Sdecompx}
\end{eqnarray}
we find, in addition to (\ref{bns}), another set of propagation eigenmodes
\begin{eqnarray}
\hat {\tilde{b}}^{in}_n&=&\int d\omega \,\psi^*_n(\omega) \hat a^{in}(\omega) \nonumber \\
\hat {\tilde{b}}^{out}_n&=&\int d\omega \,\phi^*_n(\omega) \hat a^{out}(\omega) \label{bnsx}
\end{eqnarray}
with characteristic squeezing parameters $\zeta_n$.  Assuming that all $\zeta_n$ are unequal (which is typically the
case as we show below), the Bloch-Messiah decomposition is unique, and we conclude that
$\hat{\tilde{b}}^{in}_n=\hat{b}^{in}_n$ and $\hat{\tilde{b}}^{out}_n=\hat{b}^{out}_n$, which leads to
\eeqref{conjmodes}. In the time domain, this identity corresponds to the output modes being the time reversal of the
input.

Finally, let us point out that the decomposition given in \eeqref{Sdecomp} has important implications when considering
a parametric amplifier seeded with a coherent pulse. The reason for seeding the parametric amplifier is to impose a
mode in which the majority of the output photons is generated, thus overwhelming emission to other modes whose contents
can be neglected as an unimportant background \cite{Smithey92}. \eeqref{SqOpm} indicates that the output of the
parametric amplifier will contain a pure squeezed coherent state in a single mode if the seeding coherent state
occupies one of the characteristic input modes $\phi_n(\omega)$. If the seeding pulse is in superpositon of many modes
$\phi_n(\omega)$, under the action of the parametric amplifier it will generate multiple squeezed coherent modes, in
principle each one of different intensity and squeezing parameter. The decomposition in \eeqref{Sdecomp} provides one
with characteristic pulse profiles $\phi_n(\omega)$ which are optimal for seeding the amplifier.

\section{Single pair generation regime}\label{sec:1pair}

In a general case, the equations of motion for the field operators specified by \eeqref{aprop} can be solved only by
numerical means. Before we resort to numerical methods, we will discuss in this section an approximate solution to the
problem in the weak conversion limit, when only the first order of the perturbation theory is relevant. This is indeed
the case if the nonlinear interaction length $\LNL$ is much greater that the crystal length $L$, thus keeping the squeezing
weak. This approach will give us intuitive insights into the structure of the output field, and will help us to
identify parameters relevant to its characterization in the arbitrary case.

The first order approximation to \eeqref{Bvtransform} can be obtained most easily by substituting
\begin{equation}
\hat a (\o ; z)=\exp(i k(\o)z)\, \hat{a}_{I} (\o; z)
\end{equation}
into \eeqref{aprop}, then integrating it over $z$ and retaining terms up to first order in $L/\LNL$. This procedure
yields an expression for $\hat a(\omega ; L)$. Using this result we find the approximate Green functions in the form:
\begin{eqnarray}
C(\omega,\omega') &=&\delta(\omega-\omega')e^{i k(\omega) L} \label{Eq:Comegaomega'pert} \\
S(\omega,\omega') &=&\frac{L E_p(\omega+\omega')}{\LNL E_0}
\,e^{i[k(\omega)-k(\omega')]L/2}\,\mathrm{sinc}\frac{L\Delta k}{2} \nonumber \\
& & \label{S'1st}
\end{eqnarray}
where
\begin{equation} \label{Deltak}
\Delta k=k_p(\omega+\omega')-k(\omega)-k(\omega')
\end{equation}
is the phase mismatch between the pump and the pair of down-converted photons.

In most cases it is
sufficient to expand $\Delta k$ up to the second order in deviations from the respective central frequencies. Then the
approximate expression for the phase mismatch takes the form:
\begin{eqnarray}
\Delta k&\simeq&(\beta_{1,p}-\beta_1)(\o+\o'-\o_p) \nonumber \\
&&-\frac{1}{2}\beta_2\left((\o-\o_p/2)^2+(\o'-\o_p/2)^2\right) \nonumber \\
&&+\frac{1}{2}\beta_{2,p}(\omega+\omega'-\o_p)^2,
\label{Eq:DeltakExpansion}
\end{eqnarray}
where
\begin{subequations}\begin{align}
\beta_n&=\frac{d^n k(\omega)}{d\omega^n}\Big|_{\omega=\omega_p/2} \\
\beta_{n,p}&=\frac{d^n k_p(\omega)}{d\omega^n}\Big|_{\omega=\omega_p}
\end{align}\end{subequations}
are dispersion coefficients for the pump and the signal fields. The coefficients relevant to our analysis are: the
difference of inverse group velocities $\beta_{1,p}-\beta_1$ and the group velocity dispersion coefficients $\beta_2$
and $\beta_{2,p}$. The phase term $e^{i[k(\omega)-k(\omega')]L/2}$ in \eeqref{S'1st} arises form the free propagation
of the squeezed field in the waveguide.

\begin{figure}
  \center\includegraphics[width=0.35\textwidth]{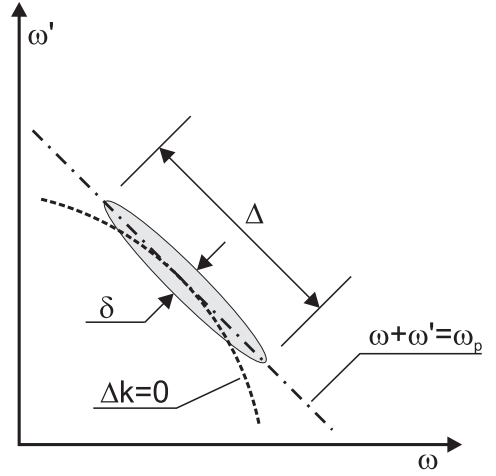}
  \caption{Gaussian approximation of the Green function $S(\omega,\omega')$, given in \eeqref{S'1st} by a product
  of the pump spectral amplitude and the phasematching function. The first factor is nonzero on a wide strip centered
  around the $\o_s+\o_i=\o_p$ line. The second factor is nonzero on a strip lying along the $\Delta k=0$ line, which is
  approximately parabolic and tangent to the line $\o_s+\o_i=\o_p$ at the point $(\o_p/2,\o_p/2)$.
  Thus the product $S(\omega,\omega')$ is nonzero on a strip ariound the $\o_s+\o_i=\o_p$ line
  and extending to points where the phasematching function goes off the pump spectral amplitude.
  }\label{sspectrum}
\end{figure}

Physically, performing the expansion up to the first order in $L/\LNL$ means considering only
generation of single photon pairs. The spectral properties of single photon pairs generated by
ultrashort pulses have been extensively studied in the context of quantum indistinguishability in
two-photon interference experiments \cite{GricePRA97,Keller,Perina,Law00}. The basic entity in
those studies is the two-photon wave function $\Psi(\omega,\omega')=\langle
1_{\omega},\,1_{\omega'}|\Psi^{out} \rangle$ describing the probability amplitude of generating a
pair of photons with frequencies $\omega$ and $\omega'$. In the Appendix, we relate this wave
function to the Green function $S(\omega,\omega')$ through a simple formula:
\begin{equation}\label{biphotonsp}
\Psi(\omega,\omega')=S(\omega,\omega')e^{i  k(\omega') L}.
\end{equation}
In other words, the eigenmodes of pulsed squeezing are closely related to the eigenmodes of the Schmidt decomposition
of a biphoton spectrum originally found by Law {\em et al.} \cite{Law00}.

Identity (\ref{biphotonsp}) suggests a Gaussian approximation similar to that applied in the analysis of the two-photon
wave function \cite{GricePRA01}, which will yield a singular value decomposition in a closed analytical form. For this
purpose, let us assume that the quadratic form of the Gaussian which approximates \eeqref{S'1st} has the principal axes
given by $\omega-\omega'=0$ and $\omega+\omega'=\omega_p$, and seek an expression of the form:
\begin{align}
S_G(\omega,\omega')=&\sqrt{\frac{2N}{\pi\delta\Delta}}\,e^{i[k(\omega)-k(\omega')]L/2} \nonumber \\
&\times\exp\left[-\frac{(\o+\o'-\o_p)^2}{2\delta^2}-\frac{(\o-\o')^2}{2\Delta^2}\right]
\label{spectrum}
\end{align}
In order to find the unknown coefficients in the above formula,
let us take a Gaussian pump pulse of duration $\tau_p$:
\begin{equation}\label{eq:pump}
E_p(\omega) \propto \exp\left[- \frac{\tau_p^2 }{2} (\omega-\omega_p)^2 \right].
\end{equation}
We insert the expansion given in \eeqref{Eq:DeltakExpansion} into \eeqref{S'1st}, apply the following approximations
to the sinc function along the axes $\omega-\omega'=0$ and $\omega+\omega'=\omega_p$:
\begin{equation}
\mathrm{sinc} \, x \approx e^{-x^2/5}, \;
\mathrm{sinc}(x^2) \approx e^{-x^2/3},
\end{equation}
and compare Eq.~(\ref{S'1st}) after all these approximations with Eq.~(\ref{spectrum}) along the principal axes.
This yields:
\begin{subequations}\label{eqs:delta}
\begin{eqnarray}
\frac{1}{\delta^2} & \approx & \tau_p^2+\frac{L^2}{10}(\beta_1-\beta_{1,p})^2 \label{eq:delta}\\
\frac{1}{\Delta^2} & \approx & \frac{1}{12}L\beta_2 \\
N & = & \int |S_G(\o,\o')|^2 d\o d\o' = \frac{L^2}{4\LNL^2}\tau_p^2\delta\Delta\quad \label{eq:defN}
\end{eqnarray}
\end{subequations}
The meaning of these parameters is depicted in Fig.~\ref{sspectrum}. Typically \cite{Shapiro}, $\delta \ll \Delta$ and
consequently the bandwidth of the downconverted light is approximately proportional to $\Delta$. In the particular
case of type I phasematching this bandwidth is limited mostly by the group velocity dispersion $\beta_2$ at signal
frequencies. Note that, in contrast, the difference between the group velocities at the pump and the signal
frequencies $\beta_1-\beta_{1,p}$ sets no limit on the downconversion spectrum. The second relevant frequency scale is
set by the parameter $\delta$ which characterizes the width of the spectral correlation between photons in pair. This
width is limited by the pump pulse bandwidth and the group velocity difference $\beta_1-\beta_{1,p}$. Also, let us
note that the Green function $S(\o,\o')$ is not normalized, and its contribution is implicitly assumed to be small, of
the order $L/\LNL$, as compared to that of $C(\o,\o')$. The normalization constant $N$ of $S_G(\o,\o')$ given in
\eeqref{eq:defN} is equal to the total number of photons per pulse emitted by the downconversion source.

Since in general $\Delta$ and $\delta$ are different, one cannot simply factorize $S(\omega,\omega')$ and define a
single mode into which photon pairs are emitted. If the signal and idler photons were nondegenerate, one would call
such a spectrum entangled; this definition is however inapplicable to the present situation because the two fields
cannot be separated using another degree of freedom. With the Gaussian approximation in hand, let us now employ the
summation formula used previously to find the Schmidt decomposition of a bipartite Gaussian state
\cite{ActaPhysicaSlovaca,GricePRA01}, which can be applied directly to the Green function (\ref{spectrum}) in order to
find its singular value decomposition. Recasting the decomposition according to the general expression specified in
\eeqref{Sdecomp}, we obtain that the
 squeezing parameters $\zeta_n$ form a
geometric sequence given by:
\begin{equation}
\zeta_n \approx \sinh \zeta_n =\frac{\sqrt{N}}{\cosh r}\tanh^{n} r \label{lambda_n}
\end{equation}
with the parameter $r=\ln(\Delta/\delta)/2$, while the corresponding eigenmodes
are described by the Hermite functions:
\begin{eqnarray}
e^{-ik(\omega)L/2}\psi_n(\omega) & = & e^{ik(\omega)L/2}\phi_n(\omega)
\label{modes1} \\
& = & \frac{H_n(\omega\tau_s)}{\sqrt{2^nn!}} \sqrt{\frac{\tau_s}{\sqrt{\pi}}}e^{-\tau_s^2(\o-\o_p/2)^2/2}, \nonumber
\end{eqnarray}
with characteristic spectral width of all squeezing modes being proportional to the inverse
geometric mean of the two spectral widths $\delta$ and $\Delta$:
\begin{equation}\label{tausq}
\tau_s=\sqrt{\frac{2}{\delta\Delta}}.
\end{equation}
These expressions for the eigenmodes and the squeezing parameters will serve as a reference when analyzing numerical
results in the nonperturbative regime presented in the next section.

As expected, $\psi_n(\omega)=\phi^*_n(\omega)$, and the modes' spectral phases can be attributed to the linear
dispersive propagation through the BBO crystal. The nonzero phase profiles are difficult to handle experimentally. One
might think they can be made negligible by choosing a sufficiently long pump pulse so its dispersion can be
neglected. This is however not the case because the eigenmodes' characteristic spectral width $1/\tau_s$ is determined
not only by $\delta$, but also by $\Delta$. The magnitude $\Delta$ is, in turn, determined by the properties of the
crystal rather than the pump pulse and is usually large.

The fact that the phase profiles of squeezing eigenmodes depend on the crystal length has an interesting consequence.
Suppose we prepare a particular state of light in a squeezing eigenmode $b_n^{in}(L)$ of a crystal of length $L$ and
insert it into the crystal. After propagation through the entire crystal, this state will undergo squeezing and leave
the crystal in the mode $b_n^{out}(L)$. Now we ask, what will be the state of light at the point $z=0$ half way
through the crystal? Propagation of light to this point is equivalent to that through a crystal of length $L/2$. The
mode $b_n^{in}(L)$ is however \emph{not} an eigenmode associated with this crystal; therefore, the state of light at
$z=0$ cannot be defined in any pure mode, but only as an entangled state of many modes. We see that while light enters
and leaves the crystal in a pure optical mode and state, inside the crystal all input modes temporarily become
entangled with each other.

\section{Intense generation regime}\label{sec:intense}

In this section, we discuss numerical solutions to the propagation equation \eqref{aprop} in a general, not
necessarily perturbative regime. As a concrete example, we will consider type-I interaction in a BBO crystal. We assume
that a waveguide is formed in a direction of perfect phase matching for an interaction of a 400~nm pump wave with
800~nm signal waves (corresponding to the angle between the waveguide axis and the optical axis $\theta=29.2^o$). For this
system, we have solved numerically \eeqref{aprop} with $k(\omega)$ and $k_p(\omega)$ approximated by values for a bulk
BBO crystal, and a Gaussian pump field given by \eeqref{eq:pump}. Let us stress that in the calculations presented
below we use neither the crude Gaussian approximation of the sinc phasematching function, nor are we limited to the
first order accuracy in $L/\LNL$. Nevertheless, the simple model developed in the preceding section will guide
us through general solutions.

As the propagation equation is linear in the strong undepleted pump
approximation, \eeqref{aprop} is identical to the classical equation
of motion for a pulse propagating through a waveguide, with the
annihilation operators $\hat a(\omega;z)$ replaced by the spectral
amplitudes of the electric field at respective signal frequencies
$\alpha(\omega;z)$ \cite{KlasycznaReferencja}. Thus we can compute
numerically the matrix approximations to the Green functions
$C(\o,\o')$ and $S(\o,\o')$ using the well-established in classical
nonlinear optics split-step method for solving partial differential
equations. The approximations are obtained by solving the
propagation equation for a complete set of $\delta$-function shaped
initial conditions where the classical signal field
$\alpha(\omega;z=-L/2)$ is taken to be equal to zero everywhere on
the computational grid in the frequency domain, except for a single
point of the grid at which it assumes $1$ or $i$. When the
distinguished point corresponds to the frequency $\o'$, solving the
propagation equation with such initial conditions yields single
columns of the discretized Green functions $C(\o,\o')$ and
$S(\o,\o')$. After calculating all the columns of the Green
functions, we computed the singular-value decompositions defined in
\eeqref{Sdecomp} which gives the desired Bloch-Messiah reduction.

We assessed whether the computational grid was fine enough to support all the modes
with nonunit gain by checking that the squeezing parameters for high-order modes approach zero. Additional tests of
computational accuracy consisted in verifying that the singular values of $C(\o,\o')$ and $S(\o,\o')$ are pairwise sinh
and cosh functions of the same real number, as stated in \eeqref{Sdecomp} and that the calculated modes uphold the
relation (\ref{conjmodes}). Although many modes are not included in our numerical calculation, we argue that their
squeezing is negligible, thus the actual choice of the mode functions $\phi_n(\o)$ and $\psi_n(\o)$ representing them
is unimportant.

Following the approach used in nonlinear optics, the calculations were performed in the reference frame of the
down-converted light, moving with the group velocity $1/\beta_1$. This corresponds to substituting
\begin{equation}\label{eq:alphapartialinteraction}
\hat{a}(\omega;z) \rightarrow e^{i[\beta_0 + \beta_1(\o-\o_p/2) ] z} \,\hat{a}(\omega;z)
\end{equation}
into the propagation equation given in \eeqref{aprop}. The mode functions $\phi_n(\omega)$ and $\psi_n(\omega)$
will be given also in the moving reference frame. This reduces to stripping the mode functions from the linear phase
in the expansion around $\omega_p/2$. Experimentally, the linear phase can be always corrected by adjusting the
temporal delay of the down-conversion beam, whereas higher order phase terms lead to pulse distortions requiring dispersion
management.

In our numerical calculations we used a Gaussian pump pulse of length $\tau_p=24$~fs (40~fs intensity FWHM) and the
waveguide length $L=1$~mm. This corresponds to the inverse bandwidth of the downconversion light $\Delta^{-1} \simeq
4$~fs and the inverse of the spectral width of correlations between the two photon in a pair $\delta^{-1}\simeq
50$~fs. We will study how the features of the down-converted light depend on the nonlinear interaction length
$\LNL$.

\begin{figure}
  \center\includegraphics[width=0.45\textwidth]{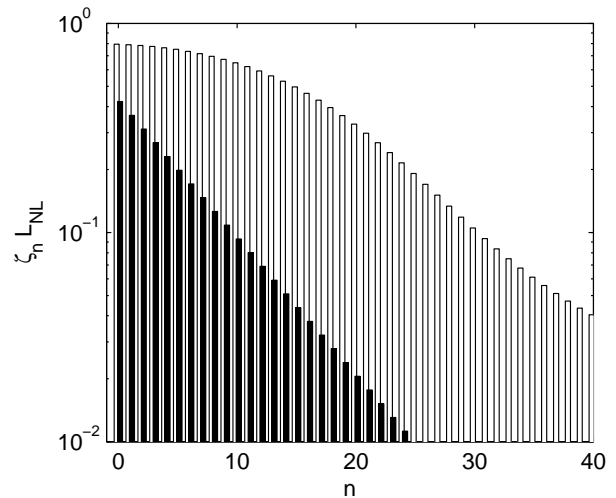}
  \caption{The rescaled squeezing parameters $\LNL\zeta_n$ as a function of the mode number, verified numerically
  to be independent of $\LNL$ for $L/\LNL \le 15$.
  Empty bars depict results of the numerical computation, while black bars represent predictions of the Gaussian approximation.}
  \label{gaind}
\end{figure}

We found that up to $L/\LNL=15$ (which corresponds to over 100~dB maximal quadrature squeezing) the \emph{squeezing
parameter} of any squeezed mode is inversely proportional to nonlinear length with a specific proportionality
constant. These values are depicted in Fig.~\ref{gaind}. We note a substantial discrepancy with the Gaussian model
from Sec.~\ref{sec:1pair}, which predicts that the squeezing parameters should follow a geometric sequence
(\ref{lambda_n}), represented by a straight line in the semi-log scale plot in Fig.~\ref{gaind}.

\begin{figure}
  \center\includegraphics[width=0.45\textwidth]{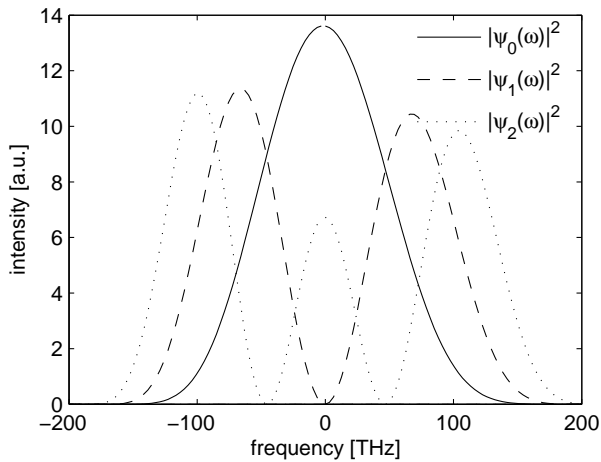}
  \caption{Spectral intensities $|\psi_n(\omega)|^2$ of the dominant three modes $n=0$ (solid line), $n=1$
  (dashed line), and $n=2$ (dotted line) for $\LNL=100$~mm.}
  \label{modes1-3}
\end{figure}

\begin{figure}
  \center\includegraphics[width=0.45\textwidth]{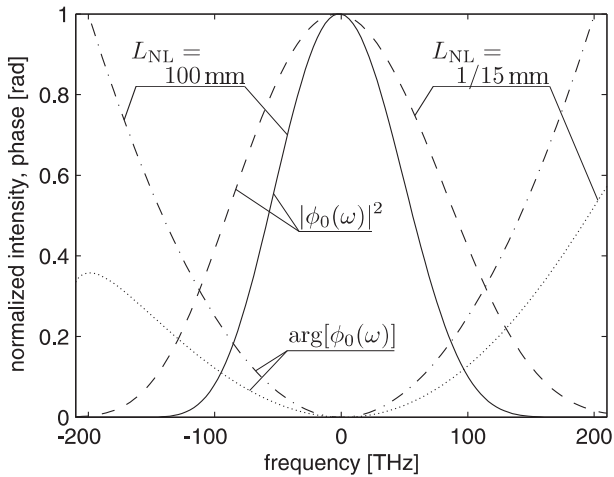}
  \caption{Spectral intensity and phase of the maximally squeezed mode $\psi_0(\omega)$
  for $\LNL=100$~mm (solid and dashed-dotted line respectively) and $\LNL=1/15$~mm (dashed and dotted lines).
  For the case studied, the input modes are complex conjugate
  to output modes, $\phi_0(\omega)=\psi_0^\ast(\omega)$.}
  \label{mode1_vs_chi}
\end{figure}

The Gaussian approximation turns out to be much more successful in predicting the \emph{shapes} of the characteristic
squeezing modes. Particularly in the weak pumping regime, the numerically calculated input and output modes
$\phi_n(\o)$ and $\psi_n(\o)$ for low $n$ are almost identical to the Hermite-Gaussian functions predicted by
\eeqref{modes1}, with only a slight asymmetry and more abrupt decay of their outermost wings (Fig.~\ref{modes1-3}).
For strong pumping, when the nonlinear interaction length $\LNL$ becomes comparable or smaller than the crystal length
$L$, the spectral intensity profiles begin to broaden and change in shape (Fig.~\ref{mode1_vs_chi}) due to nonlinear
modulation of the optical fields. Still, within the range of the interaction lengths studied here (i.e.
$L/\LNL\le 15$), the Hermite-Gaussian approximation for the several most strongly squeezed modes works reasonably
well. We expect this to be the case as long as the higher order terms in the phase mismatch expansion
(\ref{Eq:DeltakExpansion}) are negligible. In the case of BBO, this condition is satisfied for $\tau_p,\tau_s\gg
10$~fs.

\begin{figure}
  \center\includegraphics[width=0.45\textwidth]{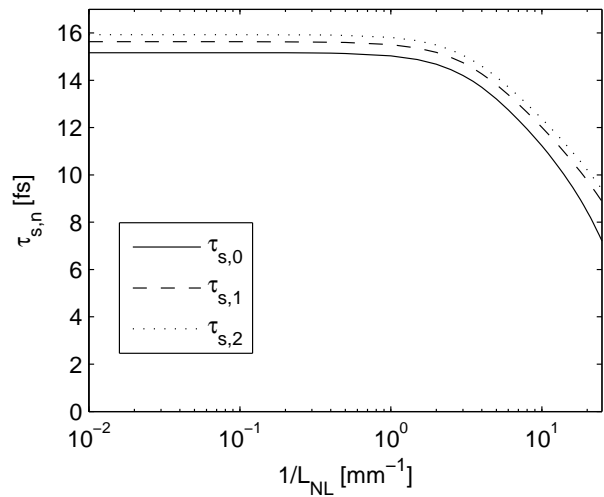}
  \caption{Duration of the dominant squeezed modes  $\tau_{s,0}$ (solid line), $\tau_{s,1}$ (dashed line) and
  $\tau_{s,2}$ (dotted line) as a function of $1/\LNL$, obtained by fitting Hermite-Gauss profiles
  from \eeqref{modes1} to numerical results.} \label{mode1tau_vs_chi}
\end{figure}

In order to characterize the change in the mode \emph{widths}, we
fitted the numerically obtained spectral intensities
$|\psi_n(\o)|^2$ with the Gaussian-Hermite approximations
\eeqref{modes1} for each $\LNL$ and $n$, treating the characteristic
time $\tau_s$ as a free parameter. We found this parameter to
deviate from the constant 20 fs value predicted by \eeqref{tausq},
and, as evidenced by Fig.~\ref{mode1tau_vs_chi}, to depend on both
the mode number and the pumping strength. In the weak pumping
regime, $\tau_s \approx 15$ fs for $n=0$ and reduces by about 3\%
for each subsequent mode, and is independent from $\LNL$. For
stronger pumping ($\LNL \gtrsim 1$), $\tau_s$ falls quickly
with the increasing pumping strength.

As expected, we found the spectral \emph{phases} of the input and the output modes to have opposite signs with
virtually equal absolute values. For strong pumping, the phase profile also  becomes flatter for shorter interaction
lengths $\LNL$, as illustrated in Fig.~\ref{mode1_vs_chi}. Thus in the intense generation regime we can no longer
attribute the spectral phase of the modes to linear dispersive propagation. All these effects mean that the features
of multimode squeezing become more fragile compared to the single-pair generation case. Thus the exact modal
characteristics need to be carefully taken into account when manipulating or detecting strongly squeezed light.

\section{Homodyne detection}\label{sec:homodyne}

Homodyne detection is a fundamental diagnostic and measurement technique in applications of squeezed states for
quantum-enhanced metrology and continuous-variable quantum information processing. It is a difficult measurement
because of the electronic noise, losses in the paths of the squeezed beams and nonideal mode matching between squeezed
field and local oscillator which defines the measured quadrature. This section is devoted to a detailed study of the
latter issue, and the modal decomposition of a parametric amplifier introduced in Sec.~\ref{sec:theory} provides a
natural framework for its discussion.

We will consider an unseeded optical parametric amplifier followed by a balanced homodyne detector with the local
oscillator (LO) pulse prepared in a certain shape characterized by the mode function $\psi_{\text{LO}}(\omega)$. The
quantity of interest is the measured quadrature noise. The analysis is straightforward in the particular case when the
local oscillator employed in the homodyne measurement is prepared in one of the characteristic modes
$\psi_{\text{LO}}(\omega)=\psi_n(\omega)$. This choice of the LO means that we are ``listening'' only to one of the
independently squeezed characteristic modes of the amplifier, whose evolution is given by \eeqref{SqOpm}. For vacuum
input, this mode evolves into a squeezed vacuum state with the squeezing parameter equal to $\zeta_n$. Thus the
detected mean square quadrature noise $\avg{Q^2_{n,\theta}}$ as a function of the LO phase $\theta$ is equal to
\cite{scaling}:
\begin{equation}
\avg{Q^2_{n,\theta}}=\frac{1}{4}(e^{2\zeta_n}\sin^2\theta+e^{-2\zeta_n}\cos^2\theta).
\label{Eq:Q2ntheta}
\end{equation}
In particular, the product of the maximum $\avg{Q_+^2}=e^{2\zeta_n}/4$ and the minimum $\avg{Q_-^2}=e^{-2\zeta_n}/4$
quadrature noise will be the same as for the vacuum state $\avg{Q_+^2}\avg{Q_-^2}=1/16$ --- the minimum value allowed
by the Heisenberg uncertainty relation.

In the general case of an arbitrary LO pulse shape we can calculate
the quadrature noise level by decomposing $\psi_{LO}(\o)$ in the basis of the characteristic output modes of the amplifier:
\begin{equation}
\psi_{LO}(\omega)=\sum_{n=0}^{\infty} M_n e^{i\theta_n} \psi_n(\omega), \label{phiLO}
\end{equation}
where $M_n$ and $\theta_n$ are real numbers and we have assumed that the spectral amplitude $\psi_{LO}(\o)$ is
normalized to unity. It is important that the quantum quadrature fluctuations of different characteristic modes are
independent because the mutually uncorrelated modes $\hat b^{in}_n$ independently evolve into $\hat b^{out}_n$. Thus,
using the local oscillator of the form (\ref{phiLO}), we measure a sum of noise contributions from the quadratures
$\avg{Q^2_{n,\theta_n}}$ weighted with $M_n^2$:
\begin{equation}\label{actualnoise}
\avg{Q^2_{LO}}=\sum_{n=0}^{\infty}\frac{M_n^2}{4}(e^{2\zeta_n}\sin^2\theta_n+e^{-2\zeta_n}\cos^2\theta_n)
\end{equation}
In general, the above expression combines contributions from both squeezed and antisqueezed quadratures. However,
if all the phases $\theta_n$ are equal, we can always bring them to $0$ by introducing an additional phase delay in the
LO arm. Then the minimum quadrature noise that can be observed is below the shot noise limit:
\begin{equation}
\avg{Q^2_-}=\sum_{n=0}^{\infty}\frac{M_n^2}{4}\exp(-2\zeta_n) \le \frac{1}{4}.
\end{equation}
The condition of equal phases $\theta_n$ can be easily realized in the single pair generation regime, because all the
output modes \eeqref{modes1} bear the same spectral phase associated with the linear propagation in the waveguide.
Thus for an unchirped local oscillator with an arbitrary intensity spectrum $I_{LO}(\omega)$ we can optimize the
amount of detected squeezing by making its phase equal to the phase acquired by the squeezed field during propagation
through a crystal of length $L/2$:
\begin{equation}\label{eq:phaselockcond}
\psi_{LO}(\omega)=\sqrt{I_{LO}(\omega)} \, e^{i k(\omega)L/2}\,.
\end{equation}

Let us note that if some of the squeezing
parameters $\zeta_n$ are equal --- for concreteness let us take $\zeta_0$ and $\zeta_1$ ---
 then any local oscillator which is in a phase locked superposition of the corresponding modes
$\psi_{LO}(\o)=M_0\psi_0(\o)+M_1\psi_1(\o)$ with real $M_0$ and $M_1$ will enable the detection of the same minimum
quadrature noise equal to $\avg{Q^2_-}=\exp(-2\zeta_0)/4=\exp(-2\zeta_1)/4$. For the weak pumping regime, several
largest $\zeta_n$ are indeed approximately (albeit not exactly) equal, as shown in Fig.~\ref{gaind}. As a result, we
have a substantial freedom in choosing the the local oscillator mode among various linear combinations of primary squeezing modes without
affecting the detection efficiency.

\begin{figure}
    \begin{center}
        \includegraphics[width=0.5\textwidth]{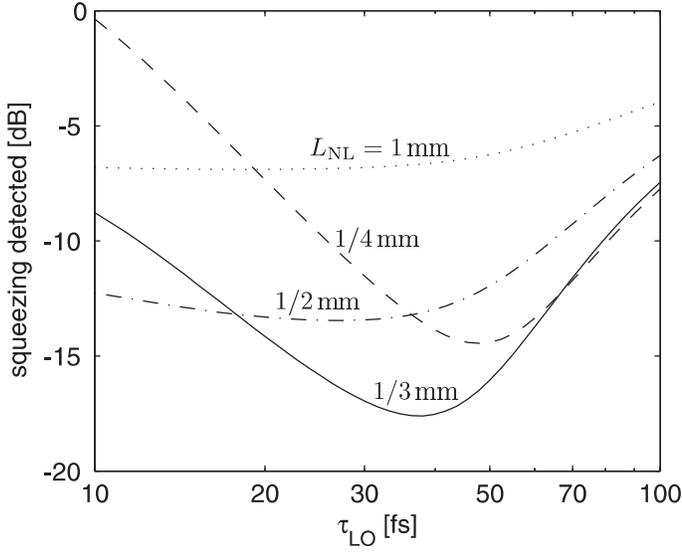}
        \caption{Minimum quadrature noise detected using the local oscillator given by \eeqref{loshape} of duration $\tau_{LO}$ for
        nonlinear lengths $\LNL=1$~mm (dotted line), $\LNL=1/2$~mm (dashed-dotted line), $\LNL=1/3$~mm (solid line),
        and $\LNL=1/4$~mm (dashed line).}
        \label{avsqu}
    \end{center}
\end{figure}

When the squeezing becomes more intense, the choice of the local oscillator mode becomes more and more critical in
order to detect squeezing. This is easily seen in the standard single-mode case described by \eeqref{Eq:Q2ntheta}, as
even a small deviation from $\theta=0$ adds a significant amount of noise from the antisqueezed quadrature. In the
case of the local oscillator distributed over several characteristic output modes of the amplifier, detection of
squeezing requires accurate control of each of the terms in its decompositon (\ref{phiLO}). This is illustrated with
Fig.~\ref{avsqu}, depicting the observed squeezing for various nonlinear interaction lengths $\LNL$, based on the
numerical results of section \ref{sec:intense}. The squeezing is calculated as a function of the inverse bandwidth
$\tau_{LO}$ of the local oscillator pulse, assumed to have a Gaussian spectral intensity profile with a phase
satisfying \eeqref{eq:phaselockcond}:
\begin{equation}\label{loshape}
\psi_{LO}(\omega)=\frac{\sqrt{\tau}}{\pi^{1/4}}\exp\left(-\frac{\omega^2\tau_{LO}^2}{2}+i\frac{L}{2}
k(\omega)\right)\,.
\end{equation}
It is seen that in the single-pair generation regime, when $L/\LNL\ll 1$, the squeezing is detected for a broad range
of local oscillator lengths. The local oscillator duration $\tau_{LO}$ can thus be swept approximately from the
inverse bandwidth of the downconversion light $\Delta^{-1} \simeq 4$~fs to the inverse of the spectral width of
correlations between the two photon in a pair $\delta^{-1}\simeq 50$~fs.

This result permits a physical interpretation of downconversion in the time-domain picture. The two photons in a pair
can be born at any moment in time within the duration of the pump pulse ``smeared'' by a difference of the group
velocities between pump and signal pulses $\beta_{1,p}-\beta_1$ [see \eeqref{eq:delta}], i.e. within the time interval
$1/\delta$. If the LO pulse of a longer duration is chosen, it is known \emph{a priori} to contain modes into which no photons have
been emitted, thus entailing inefficient detection. The existence of the upper limit $\Delta$ to the local oscillator
bandwidth is explained by the finite length of the down-conversion crystal. Due to the group velocity dispersion (which limits the down-conversion spectrum), single-photon wavepackets propagating through the crystal
diverge in time by $\Delta^{-1}$. If the local oscillator pulse is chosen too short, it may happen that one photon in
a pair is registered within the local oscillator mode, but the other one arrives either sooner or later, also leading
to a reduction in detection efficiency.

\begin{figure}
    \begin{center}
        \includegraphics[width=0.5\textwidth]{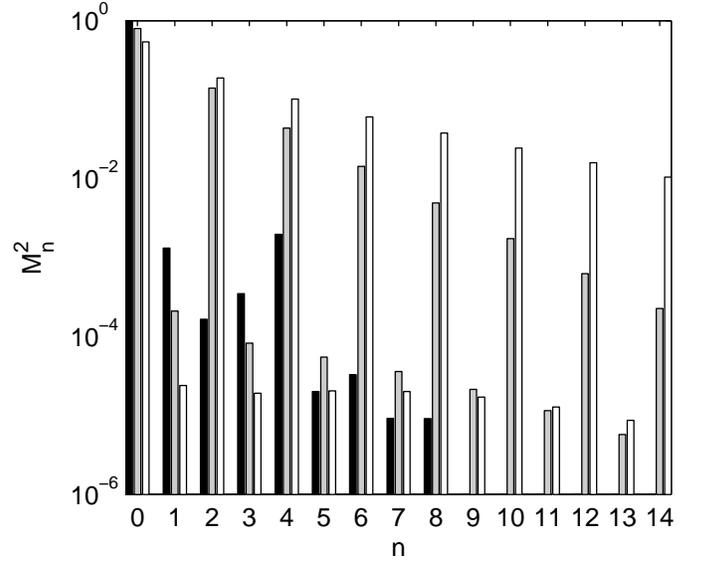}
        \caption{
        Squared coefficients $M_n^2$ for the decomposition of a local oscillator pulse given by \eeqref{loshape}
        into characteristic squeezing modes according to \eeqref{phiLO}, for the nonlinear length $\LNL=1/2$~mm
        and $\tau_\mathrm{LO}=15$~fs (black bars), $\tau_\mathrm{LO}=30$~fs (grey bars), and
        $\tau_\mathrm{LO}=50$~fs (empty bars). Note the logarithmic scale on the vertical axis.}
        \label{Mn}
    \end{center}
\end{figure}

As seen in Fig.~\ref{avsqu}, when the pumping intensity increases, reducing $\LNL$, the choice of the LO duration
$\tau_{LO}$ becomes more and more critical and finally, due to the mismatched spectral phase, no choice of $\tau_{LO}$
matches the most strongly squeezed characteristic mode. Moreover, it turns out that using an LO pulse significantly
longer than the fundamental squeezed mode is favorable. This effect appears paradoxical, because the squeezing modes'
characteristic time in fact reduces with stronger pumping (Fig.~\ref{mode1tau_vs_chi}), but it can be explained as
follows. Numerical calculations show that for $\tau_{LO}\simeq\tau_s$ the decomposition \eqref{phiLO} contains about a
2\% contribution from modes with odd $n$, as shown in Fig.~\ref{Mn}. We found that for small $n$, corresponding to the
most strongly squeezed modes, the phases in the decomposition in \eeqref{phiLO} follow approximately the rule
$\theta_n\simeq n\pi/2$. Consequently the modes with odd $n$, despite having only a minor contribution to LO shape,
build up altogether significant noise, as we are observing their noisy quadratures, described in \eeqref{actualnoise}
by terms proportional to $e^{\zeta_n}$. For $\tau_{LO}\simeq 2\tau_s$, on the other hand, even though many more modes
contribute to the LO decomposition in \eeqref{phiLO}, they are predominantly even with equal phases $\theta_n\simeq0$.
According to our simulations, the loss of squeezing at high pumping intensities is present valid even if we allow more
general LO pulse shapes with a Gaussian spectral profile and an arbitrary quadratic phase.

The quadrature noise observed in homodyne detection of multimode squeezed light is similar to that
of a single-mode squeezed state with an inefficient detector in that it does not exhibit minimum
uncertainty: $\langle Q_+^2\rangle\langle Q_-^2\rangle>1/16$. In the latter case, the efficiency of
the detector can be calculated as follows:
\begin{equation}
\eta = \frac {-16\langle Q_+^2\rangle\langle Q_-^2\rangle+4\langle Q_+^2\rangle + 4\langle
Q_-^2\rangle-1}{4\langle Q_+^2\rangle+4\langle Q_-^2\rangle-2}. \label{eta}
\end{equation}

\begin{figure}
    \begin{center}
        \includegraphics[width=0.5\textwidth]{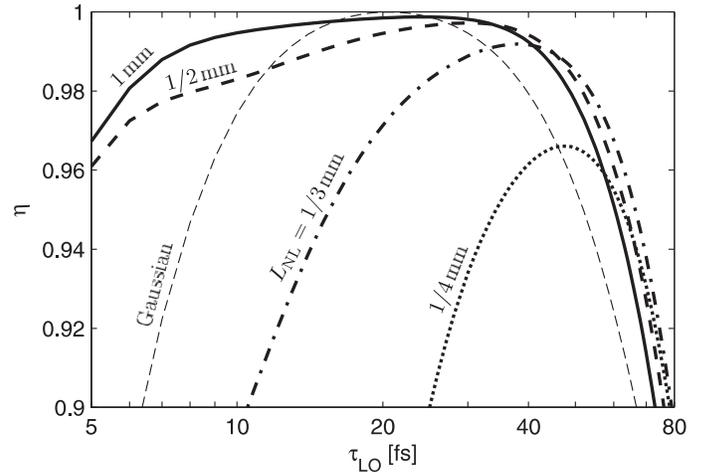}
        \caption{Quantum efficiency $\eta$ of detecting squeezing defined in \eeqref{eta} as
        a function of the local oscillator duration $\tau_{LO}$ for
        nonlinear lengths $\LNL=1$~mm (solid line), $\LNL=1/2$~mm (dashed line), $\LNL=1/3$~mm (dashed-dotted line),
        and $\LNL=1/4$~mm (dotted line). The thin dashed line represents the result of the Gaussian approximation.}
        \label{qe}
    \end{center}
\end{figure}

It is sometimes useful to express the quality of multimode squeezed light in terms of the efficiency calculated
according to the above relation. In Fig.~\ref{qe}, we plot this quantity for a local oscillator of the shape given by
\eeqref{loshape}, as a function of the LO pulse duration, for several squeezing strengths. We observe the same general
behavior as in Fig.~\ref{avsqu}. In the single-pair generation regime, high-$\eta$ squeezing detection is possible in
a broad span of LO modes within the interval $1/\Delta \le \tau_{LO} \le 1/\delta$;
as the pumping intensity is increased, the choice of LO becomes a lot more critical.

Most existing experiments on pulsed squeezing derive the local oscillator from the master laser whose second harmonic
serves as a pump for the down-conversion. According to Figs.~\ref{avsqu} and~\ref{qe}, this is not always the best
strategy. In the weak pumping regime, the highest efficiency is achieved for the LO pulse duration of $\tau_s$, which
is usually much shorter than the fundamental pulse width (\ref{tausq}). With increasing squeezing strength, the
optimal LO pulse width increases.

Combining these considerations allows one to match the parameters of the setup, such as the duration of the
fundamental and second-harmonic pulses, as well as the characteristics of the down-conversion crystal, in order to
reach the optimum detection of squeezing.

\section{Conclusions}\label{sec:conclusions}
In summary, we have analyzed the spectral properties of pulsed squeezed light generated by means of
single-pass optical parametric amplifier. The main tool in our analysis was the Bloch-Messiah
decomposition of the Green functions describing the parametric down-conversion process. This
decomposition allowed us to identify the characteristic squeezing modes, which provide a simple way
of describing quantum statistical properties of the multimode output field.

We have developed an analytical model describing the perturbative regime of single pair generation
$L/\LNL \ll 1$, when the nonlinear interaction length $\LNL$ substantially exceeds the crystal
length $L$. The model yielded an approximate form of the squeezing modes as well as their
characteristic time scale. In this regime, high-efficiency squeezing expands over a broad range of
pulsed modes temporally synchronized with the pump pulse. Some of the results of the analytical
model, such as the mode shapes, remain qualitatively valid also beyond the perturbative limit.

For the multiple-pair generation case we have presented realistic numerical calculations of the characteristic modes
of a parametric amplifier, in which pure squeezing is present. Beyond the perturbative regime, the mode functions
depend on the pumping intensity, including the change in their spectral phases. As with increasing parametric gain the
fluctuations in antisqueezed quadratures grow exponentially, it becomes more difficult to eliminate their parasitic
contribution to the observed homodyne noise. Therefore, one needs precise tailoring of the local oscillator mode,
including both its amplitude and its phase, if substantial squeezing is to be observed.

The multimode character of the output field generated by an optical parametric amplifier has important consequences
when considering quantum information applications based on continuous variables, such as quantum cryptography. For
example, if a quantum communication protocol involves modulating squeezed light, care must be taken to ensure that
only the mode detected at the receiving station experiences modulation. Otherwise other modes might carry the same
signal and could be detected by a third party without causing any observable disturbance. Furthermore, if the
quadrature fluctuations detected by the legitimate user remain above the minimum uncertainty limit, this implies that
the observed mode was correlated with other undetected modes which again could be used to gain information by a third
party. The multimode nature of pulsed squeezing also needs to be taken into account is quantum state engineering
combining discrete- and continuous-variable methods, such as preparation of squeezed single-photon states
\cite{Sasaki05,gran04}, Schr\"odinger cats \cite{lund04}, and entanglement purification \cite{cventpur}. When multimode
squeezing is undesired, it can be prevented by choosing down-conversion configuration with unentangled biphoton
spectrum \cite{GricePRA01,ure05,microcavity}.

Let us also note that the decomposition of the output of the parametric process into characteristic
squeezing modes enables a straightforward analysis of the result of combining two squeezed beams on
a balanced beam splitter. It is well known that in the single-mode picture, such a procedure yields
a pair of twin beams \cite{Ou92}. In the realistic case, if the two incident squeezed beams are
generated by identical parametric amplifies, they will comprise the same characteristic squeezing
modes perfectly matched pair-wise. Consequently, combining them on a beam splitter will yield
multimode twin beam sets which exhibit minimum-uncertainty quadrature correlations in the
characteristic modes of the squeezers.

\section*{Acknowledgements}

We acknowledge helpful discussions with M. G. Raymer, I. A.
Walmsley, and K.-P. Marzlin, as well as financial support from CFI,
NSERC, CIAR, AIF, MNiI grant number 2P03B 029 26, and the European Commission
(QAP).

\appendix

\section{Analysis in the interaction picture}
\label{Appendix:Psi}
The evolution of the quantum state of light while propagating through the crystal follows:
\begin{equation}\label{Schroedinger}
\ket{\psi_I^{out}}={\cal T} \exp\left[-i\int\limits_{-L/2}^{L/2}\hat
V_I(z) dz\right]\ket{\psi_I^{in}},
\end{equation}
where the subscript $I$ denotes the interaction picture and ${\cal T}$ denotes $z$-ordering, analogous to
standard time ordering. The three-wave mixing Hamiltonian (\ref{Ham}) can be rewritten in the interaction picture as
\begin{eqnarray}
\lefteqn{\hat V_I(z) =  \frac{1}{2\LNL E_0} \int d\omega \, d\omega'}  \label{Hamint}
&&\\
&&\times [i e^{i \Delta kz}E_p(\omega'+\omega)\hat a^\dagger(\omega)\hat a^\dagger(\omega')+{\rm H.c.}],
\nonumber
\end{eqnarray}
where $\Delta k$ is given by \eeqref{Deltak}.
In the weak conversion limit, when only the first order of the perturbation theory is relevant, we can neglect the $z$-ordering and obtain
\begin{equation}
\ket{\psi_I^{out}}=e^{\frac{1}{2}\int d\omega d\omega'\Psi_I(\omega,\omega') [\hat a^\dagger(\omega)\hat
a^\dagger(\omega') -\hat a(\omega)\hat a(\omega')] }\ket{\psi_I^{in}}\label{evolint},
\end{equation}
where the integration kernel
\begin{equation}
\Psi_I(\omega,\omega') =\frac{L E_p(\omega+\omega')}{\LNL E_0}
\,\,\mathrm{sinc}\frac{L\Delta k}{2} \nonumber \\
\label{Psi'1st}
\end{equation}
determines the frequency correlations of photons in a down-converted
pair. In the weak excitation limit with the vacuum input, the output state takes
the form:
\begin{equation}
\ket{\psi_I^{out}}=\ket{0}+\frac{1}{2}\int d\omega d\omega'\Psi_I(\omega,\omega')
\ket{1_\omega,1_{\omega'}}\label{evolintweak}.
\end{equation}

 In the Schr\"odinger picture, the output state
 acquires an additional optical phase:
\begin{equation}\label{intschro}
\Psi_S(\omega,\omega')=\Psi_I(\omega,\omega')e^{i[k(\omega)+k(\omega')]L/2}.
\end{equation}
We recover \eeqref{biphotonsp} from Eqs.~(\ref{S'1st}), (\ref{Psi'1st}), and (\ref{intschro}).

Note that the biphoton spectrum can be diagonalized according to \eeqref{Bvtransformx}
\begin{equation}
\Psi_S(\omega,\omega')=\sum\limits_{n=0}^\infty\zeta_n\psi_n(\omega)\psi_n(\omega').\label{Schmidt}
\end{equation}
and the output state can be rewritten as follows:
\begin{equation}
\ket{\psi_S^{out}}=\ket{0}+\frac{1}{2}\sum\limits_{n=0}^\infty \zeta_n(\hat b_n^{out\,\dagger})^2\ket{0}
\label{psioutbbschro},
\end{equation}
where the $\hat b_n$'s represent individual squeezing eigenmodes
(\ref{bns}). As we see, the problem of finding the characreristic
modes of a squeezer in the weak pumping limit is equivalent to
decomposing the biphoton spectrum into uncorrelated components.


\end{document}